# On the origin of nonreciprocal charge transport in topological insulator $BiSbTeSe_2$

Fengyi Guo[1], Ziqi Wang[1], Xiubing Li[1], Zheng Dai[1], Congcong Li[1], Yufan Ju[1], Shuai Zhang[1,2,3,*], Bin Cheng [1,4], Zhe Ying[3], Fengqi Song[1,2,3,*]

[1] National Laboratory of Solid State Microstructures, Collaborative Innovation Center of Advanced Microstructures, Jiangsu Physical Science Research Center, and School of Physics, Nanjing University, Nanjing 210093, China

[2] Institute of Atom Manufacturing, Nanjing University, Nanjing 211800, China

[3] Jiangsu Provincial Key Laboratory of Atomic Level Manufacturing, Nanjing Institute of Atomic Scale Manufacturing, Nanjing 211800, China

[4] School of Physics and Electronic Information, Jiangsu Second Normal University, Nanjing 211200, China

*Corresponding authors. S.Z. (szhang@nju.edu.cn), F.S. (songfengqi@nju.edu.cn)

**Abstract**

Nonreciprocal charge transport provides an effective approach for probing direction-dependent charge dynamics and spin–momentum-locked electronic states. Here, we report highly gate-tunable nonreciprocal charge transport in dual-gated $BiSbTeSe_2$ topological insulator device. The second-harmonic longitudinal resistance is linear in both current and magnetic field. The nonreciprocal response reaches its maximum near the charge neutrality point, exhibits sign reversals on either side, and is strongly suppressed once the Fermi level enters the bulk states. We suggest that the nonreciprocal response is attributed to the inhomogeneity of spin–momentum locking in the topological surface states. The nonreciprocal signal near the charge neutrality point is about one order of magnitude larger than that without gate modulation and remains robust up to 300 K, highlighting $BiSbTeSe_2$ as a promising platform for room-temperature nonlinear electronics.

**Main text**

Symmetry plays a fundamental role in determining the electronic structure and transport properties of condensed-matter systems. High symmetries impose stringent constraints on the dispersion of low-energy quasiparticles[1,2], whereas the breaking of spatial inversion, mirror, and other crystalline symmetries enables unconventional transport phenomena beyond the conventional linear-response regime[3-9]. In the linear-response regime, the longitudinal voltage is proportional to the applied current, so reversing the current merely reverses the voltage sign, leaving the resistance unaffected, such that $R(I) = R(-I)$. Likewise, in nonmagnetic systems preserving time-reversal symmetry, the longitudinal resistance generally obeys the Onsager reciprocity relations[10]. However, when spatial inversion symmetry is broken and time-reversal symmetry is further lifted by an external magnetic field or magnetic order, charge transport along opposite directions can become inequivalent. Consequently, the system may exhibit nonreciprocal charge transport (NCT) in the longitudinal channel, in which opposite current directions correspond to different resistances[6,7,11-35], i.e. $R(I) \neq R(-I)$. Such nonlinear longitudinal responses can be observed in non-centrosymmetric quantum materials. Representative material platforms for NCT[6,7,11-35] include Rashba spin-split systems, magnetic and chiral materials, superconductors, and topological surface states (TSSs). Superconducting systems exhibit prominent nonreciprocal phenomena such as the superconducting diode effect and current-induced nonreciprocal transport[36-39]. In TSSs, bilinear magnetoelectric resistance (BMER) provides a representative example of nonreciprocal charge transport [11,12,20]. The TSSs constitute a spin-nondegenerate system arising from surface inversion-symmetry breaking combined with strong spin–orbit coupling[40,41]. The nonlinear longitudinal resistance can be phenomenologically expressed as $R_{xx}^{2\omega} = \gamma R_0 \boldsymbol{I} \cdot (\boldsymbol{P} \times \boldsymbol{B})$, where $R_0$ is the resistance at zero magnetic field, $\boldsymbol{I}$ is the electric current, $\boldsymbol{B}$ is the magnetic field, and $\boldsymbol{P}$ represents the polar direction of the conductor[14,33]. The magnitude of the nonreciprocal transport is determined by the nonreciprocal coefficient $\gamma$.

In topological insulators (TIs), the origin of nonreciprocal charge transport has been attributed to different mechanisms[11,12,20]. In $Bi_2Se_3$[12,20,42], the nonlinear longitudinal response is found to be associated with the hexagonally warped Fermi surface of the spin-momentum-locked TSSs. In contrast, nonreciprocal charge transport in strained HgTe[11], where hexagonal warping is absent[33], is attributed to the spin-momentum-locking inhomogeneity. However, relatively few studies have explored gate control in TIs with warped Fermi surfaces, and the use of dual-gate configurations to investigate the underlying scattering mechanisms remains limited[14]. Consequently, the respective roles of these two mechanisms in nonreciprocal charge transport remain a subject of ongoing debate. Moreover, the temperature evolution of the nonreciprocal charge response, particularly whether it can persist over a broad temperature range and remain robust up to room temperature, has yet to be investigated. As a representative Bi-based three-dimensional TI, $BiSbTeSe_2$ (BSTS) is bulk-insulating, which not only allows transport to be dominated by the TSSs but also permits effective tuning of its Dirac carriers through electrostatic gating[43-45]. Therefore, dual-gated BSTS device provides an ideal platform for investigating the nonreciprocal charge transport.

In this work, we observe the highly gate-tunable nonreciprocal charge transport in dual-gated BSTS flake device. A pronounced maximin magnitude is observed when the Fermi level ($E_F$) approaches the charge neutrality point (CNP), followed by a sign reversal as the $E_F$ is further tuned across the CNP. It indicates that the nonreciprocal response in BSTS near the CNP originates from scattering associated with inhomogeneous spin–momentum locking. Furthermore, the magnetic-field angular dependent behavior also suggests that the contribution from hexagonal warping is not dominant. The nonreciprocal coefficient $\gamma$ in BSTS is approximately two orders of magnitude larger than that in $Bi_2Se_3$, and the nonreciprocal response can persist up to 300 K. Our work provides deep insight into the nonreciprocal charge transport of TSSs and demonstrates the potential of BSTS for room-temperature nonreciprocal electronic device.

The thin $BiSbTeSe_2$ flake was mechanically exfoliated onto Si substrates with 300-nm-thick $SiO_2$. Gold electrodes were fabricated by electron beam lithography and electron beam evaporation, followed by dry transfer of hexagonal boron nitride ($h$-BN) and graphite onto the BSTS flake. The thickness of BSTS is about 40.7nm (see Fig. S1 in the Supplementary Information). Figure 1(a) shows the transport measurement configuration, where graphite/$h$-BN and $SiO_2$/Si serve as the top gate ($V_{tg}$) and back gate ($V_{bg}$), respectively. The coordinate system and the definitions of the magnetic-field rotation angles $\theta$, $\alpha$, and $\varphi$ are indicated**.** The transport was measured by standard lock-in amplifiers (SRS SR830). An excitation current with a low frequency of 31 Hz was applied. In Fig. 1(b), as the temperature ($T$) decreases, the longitudinal resistance $R_{xx}$ first increases and then decreases, indicating that the TSSs dominate the transport at low temperatures. Figure 1(c) presents the dual-gate mapping of the longitudinal resistance $R_{xx}$ at $T = 20$ K without external magnetic field, and the corresponding linecut at a fixed back-gate voltage $V_{\mathrm{bg}} = 2$ V is shown in Fig. 1(d). Unlike the cross-shaped pattern reported in dual-gated TI device[46,47], our mapping shows only a single diagonal resistance ridge, indicating that the top and bottom surfaces cannot be independently controlled by dual gates. The resistance exhibits a pronounced maximum at ($V_{\mathrm{tg}}$, $V_{\mathrm{bg}}$) = (-2 V, 2 V), identifying the CNP of the BSTS device. Additional basic electrical transport characterizations are presented in Fig. S2 in the Supplementary Information.

Then, the nonreciprocal charge transport of TSSs is studied. Figure 2a shows the magnetic-field-dependent nonreciprocal resistance $R_{xx}^{2\omega}$ ($R_{xx}^{2\omega} = V_{xx}^{2\omega}/I^{\omega}$.) of the BSTS at 20 K under different driving currents, with the top- and back-gate voltages fixed near the CNP. The data have been antisymmetrized with respect to the magnetic field. For all currents, $R_{xx}^{2\omega}$ varies linearly with magnetic fields. The extracted current dependence of $R_{xx}^{2\omega}$ at $B_{\mathrm{y}} = 1$ T is displayed in Fig. 2(b). The current dependence of $R_{xx}^{2\omega}$ remains linear up to $I^{\omega} = 5$ μA. Thus, the nonreciprocal resistance $R_{xx}^{2\omega}$ scales linearly with the driving current and magnetic field, satisfying the requirements for nonreciprocal charge transport. In Fig. 2(c), the second-harmonic voltage $V_{xx}^{2\omega}$ as a function of current overlaps well under different frequencies, ruling out the possible spurious capacitive coupling effect[34]. The measurements with reversed current and magnetic-field directions, as well as different contact configurations, are provided in Fig. S3 in the Supplementary Information.

The gate-dependent nonreciprocal charge transport is shown in Fig. 3. Figure 3(a) shows the antisymmetrized dual-gate mapping of the $R_{xx}^{2\omega}$ at $B_y$ = 1 T, using the same gate-voltage range as in Fig. 1(e). In the dual-gate mapping, the strongest nonreciprocal resistance occurs in the central dark-blue region corresponding to the CNP, while the light-red region off the CNP shows a reversed sign and a weaker magnitude. The corresponding line cut of $R_{xx}^{2\omega}$ and the nonreciprocal coefficient $\gamma = \frac{R_{xx}^{2\omega}}{R_0 IB}$ at $V_{bg}$ = 2 V are shown in Figs. 3(b) and 3(c), respectively. Both $R_{xx}^{2\omega}$ and $\gamma$ reach their maximum amplitude near the CNP, and undergo sign reversals at $V_{tg}$ = -3.8 V and $V_{tg}$ = 0.2 V. In contrast, $R_{xx}^{2\omega}$ and $\gamma$ are substantially reduced in the region far away from the CNP. When $E_F$ is tuned far from the CNP and enters the bulk states, the nonreciprocal response becomes nearly negligible, because the bulk states preserve inversion symmetry and therefore do not generate a nonreciprocal response.

The pronounced nonreciprocal response near the CNP indicates that nonreciprocal transport in BSTS is closely related to its band structure and strongly depends on the $E_F$. Two mechanisms have been proposed for nonreciprocal charge transport in TIs. The first originates from the second-order response of a warped Fermi surface. Accordingly, the nonreciprocal response amplitude is proportional to the square of the hexagonal warping strength and increases linearly with $|\varepsilon_F|$, where $\varepsilon_F = E_F - E_{CNP}$ represents the energy position of the Fermi level relative to the CNP. The second mechanism relies on the dependence of the scattering rate on the current-induced spin polarization, for which the magnitude of the nonreciprocal response[14] is proportional to $|\varepsilon_F|^{-3}$.

Here, the largest magnitude of the nonreciprocal response in BSTS is observed near the CNP, which suggests that the nonreciprocal charge transport is attributed to scattering arising from inhomogeneities of spin–momentum locking in the TSSs. Moreover, the sign reversals on both sides of the CNP provide further support for this interpretation. We can decompose the total nonreciprocal resistance into separate contributions from the top and bottom surfaces, $R_{xx}^{2\omega} = R_{xx_t}^{2\omega} + R_{xx_b}^{2\omega}$. The top (t) and bottom (b) surfaces of a TI host TSSs with opposite chirality, hence, the nonreciprocal resistance have opposite signs at the top and bottom surfaces. In TIs, the nonreciprocal response of the two-surface system is governed by the competition between the top and bottom surface channels[14]. By considering the current partition between the two surfaces, with $I_{t(b)} = I_{TSS}\frac{\sigma_{t(b)}}{\sigma_t + \sigma_b}$ and $\sigma_{t(b)} \propto \left(\varepsilon_{F_t(b)}\right)^2$, the total nonreciprocal response can be written as $\gamma \propto I_{TSS} \times \frac{|\varepsilon_{F_b}| - |\varepsilon_{F_t}|}{\left(\varepsilon_{F_b}\right)^2 + \left(\varepsilon_{F_t}\right)^2}$, where $\varepsilon_{F_b(t)} = E_F - E_{CNP_b(t)}$. This expression shows that the nonreciprocal response is governed by the imbalance between the top and bottom surface contributions. Near the CNP, the $\left(\varepsilon_{F_t(b)}\right)^2$ are very small, strongly amplifying the nonreciprocal coefficient. This accounts for the maximum magnitude of $\gamma$ near the CNP. When the gate voltage is tuned away from the CNP the denominator is no longer strongly suppressed, and the amplification effect becomes weaker. When $\varepsilon_{F_t}$ equals to $\varepsilon_{F_b}$, the contributions from top and bottom surface cancel each other and the nonreciprocal response crosses zero (see Figs. S4 and S5 in the Supplementary Information for detailed discussion).

In addition, previous studies have demonstrated that the angular dependence of nonreciprocal response can be used to probe the spin texture of TSSs, with the out-of-plane spin component associated with the hexagonal warping[12]. We therefore measured the angular-dependent nonreciprocal response by rotating the magnetic field in the *xy, xz,* and *yz* planes at $T = 20$ K and $|B| = 1$ T without gate voltages, as shown in Fig. 4 (also see Fig. S6 in the Supplementary Information). The longitudinal resistance $R_{xx}$ (red curves) exhibits the expected 180° periodicity in all three rotation planes. The nonreciprocal resistance $R_{xx}^{2\omega}$ (blue curves) shows a 360° periodicity in the *xy* and *yz* planes, while it is strongly suppressed in the *xz* plane, indicating a polar axis predominantly along the z direction, i.e., $\boldsymbol{P_z}$. In particular, previous works have observed a finite and clearly response in the *xz* plane, with a magnitude that can be directly compared with the responses in other rotation planes. Such an *xz*-plane response is expected to be sensitive to the out-of-plane spin component induced by hexagonal warping. In contrast, the nonreciprocal signal in the *xz* plane is almost zero in the BSTS device, without a discernible characteristic angular dependence. Therefore, the hexagonal warping is unlikely to be the dominant origin of the nonreciprocal response in our work.

Then, we tuned the $E_F$ close to the CNP and investigated the temperature evolution of nonreciprocal charge transport. Figure 5(a) show the magnetic-field dependence of the nonreciprocal resistance $R_{xx}^{2\omega}$ at different temperatures near the CNP. Compared with the response without gate modulation in Fig. 5(b), where $E_F$ is far from the CNP, the nonreciprocal response near the CNP is enhanced by approximately one order of magnitude and remains robust up to 300 K. These results demonstrate that electrostatic gating can effectively tune and significantly enhance the nonreciprocal charge transport in BSTS. The evolution of $\gamma$ with temperature is further compared in Fig. 5(c) for the CNP-tuned state. Notably, a rapid enhancement of the nonreciprocal coefficient $\gamma$ is found upon cooling below approximately 75 K, which coincides with the temperature range where the transport becomes dominated by the TSSs. At higher temperatures, bulk carriers participate in charge transport but do not contribute to the nonreciprocal response, which suppresses the nonreciprocal coefficient $\gamma$. This correlation indicates that the enhancement of nonreciprocal charge transport is closely linked to the onset of TSSs-dominated transport (see Fig. S7 in the Supplementary Information for detailed discussion).

Then, we compare the nonreciprocal transport coefficient $\gamma$ of BSTS with that of other TIs. As summarized in Table 1, the value of $\gamma$ obtained in BSTS device is approximately two orders of magnitude larger than that previously reported for $Bi_2Se_3$, demonstrating the highly efficient nonreciprocal transport in BSTS. This enhancement can be attributed to the more pronounced transport contribution from the TSSs near the CNP than in regions far away from the CNP, which makes the scattering effect associated with spin–momentum locking more prominent. These results suggest that scattering associated with spin-momentum locking plays a dominant role in generating nonreciprocal charge transport in TIs.

In summary, we have observed pronounced nonreciprocal charge transport in dual-gated BSTS device and realized effective gate control over both the magnitude and sign of the nonreciprocal response. The nonreciprocal signal is significantly enhanced near the CNP, indicating that the response originates from the scattering processes associated with the spin–momentum locking inhomogeneity of the TSSs. Notably, the nonreciprocal charge response remains robust up to room temperature, and the corresponding nonreciprocal coefficient $\gamma$ is large, demonstrating the high efficiency of nonreciprocal charge transport in BSTS. These results highlight the potential of BSTS for room-temperature nonlinear electronics and nonreciprocal device applications.

**ACKNOWLEDGEMENTS**

We acknowledge the support of the National Natural Science Foundation of China (Grant Nos. 12374043, 92580203, 12504053, T2221003), the National Key R&D Program of China (No. 2022YFA1402404), the Natural Science Foundation of Jiangsu Province (Nos. BK20240166, BK20243013, BK20233001), the Fundamental and Interdisciplinary Disciplines Breakthrough Plan of the Ministry of Education of China (No. JYB2025XDXM411), and the Fundamental Research Funds for the Central Universities (Nos. KG202501, KG202603).

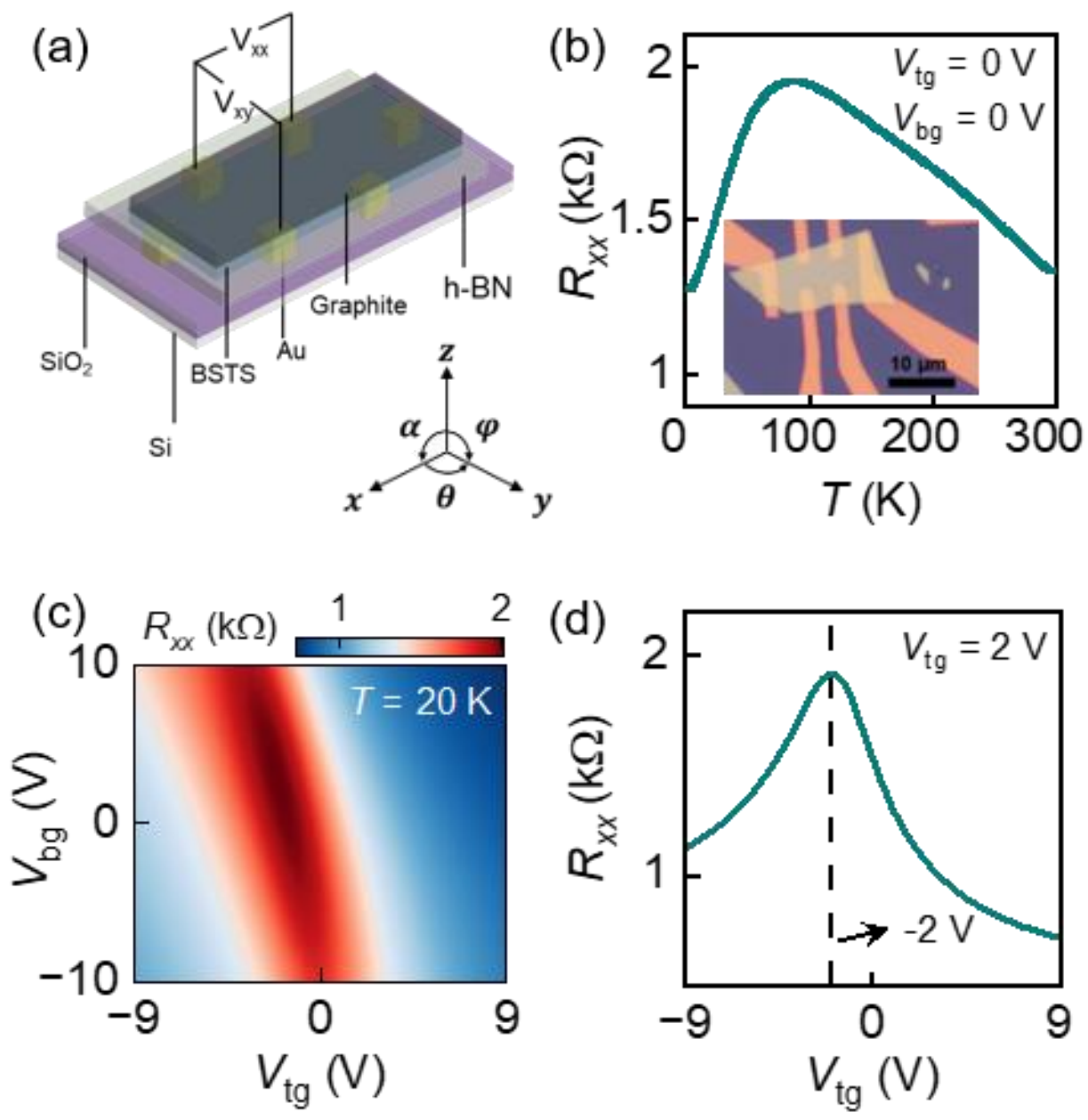

**FIG. 1**. (a) The structure of the dual-gated BSTS device. Si/$SiO_2$ and graphite/*h*-BN work as back and top gate, respectively. (b) Temperature dependent resistance $R_{xx}$. Inset is the optical image of the BSTS device. (c) The mapping of longitudinal resistance $R_{xx}$ as a function of top gate voltage ($V_{tg}$) and bottom gate voltage ($V_{bg}$) at 20 K. (d) The $V_{tg}$ dependent $R_{xx}$ at $V_{bg}$ = 2 V.

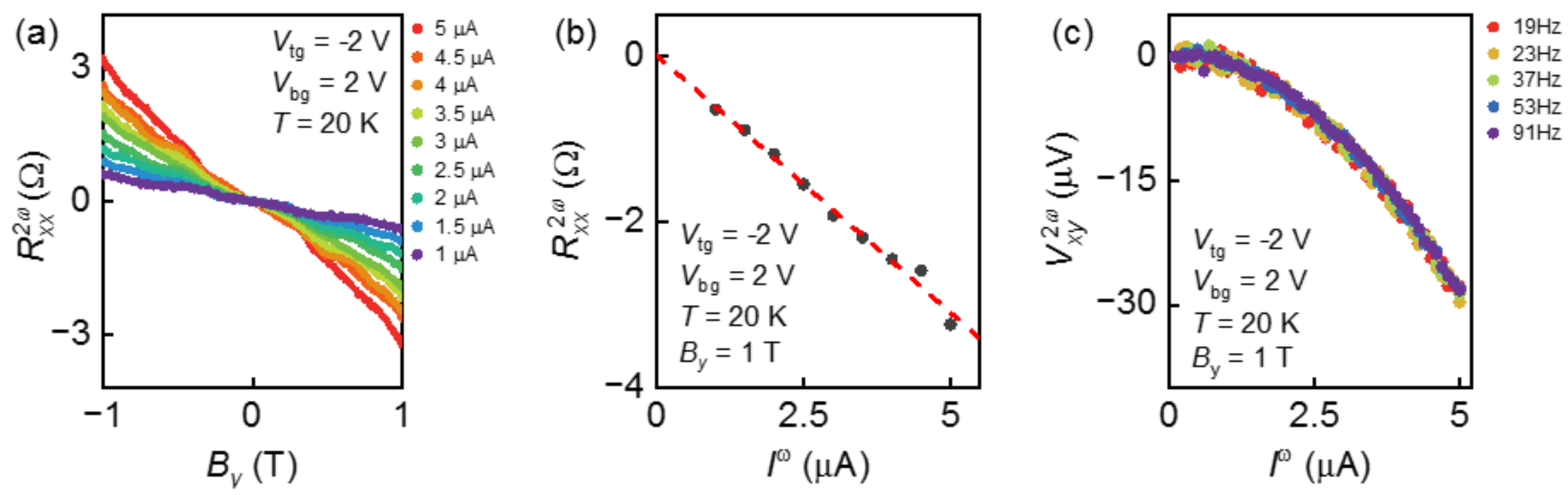

**FIG. 2**. (a) Magnetic field dependence of $R_{xx}^{2\omega}$ under different currents measured at 20 K, with the top- and back-gate voltages fixed at ($V_{\rm tg}$, $V_{\rm bg}$) = (-2 V, 2 V). (b) Current dependence of $R_{xx}^{2\omega}$ at $B_{\rm y}$ = 1 T. $R_{xx}^{2\omega}$ exhibits a linear dependence on the current. (c) Current dependent second-harmonic longitudinal voltages $V_{xx}^{2\omega}$ under different frequencies.

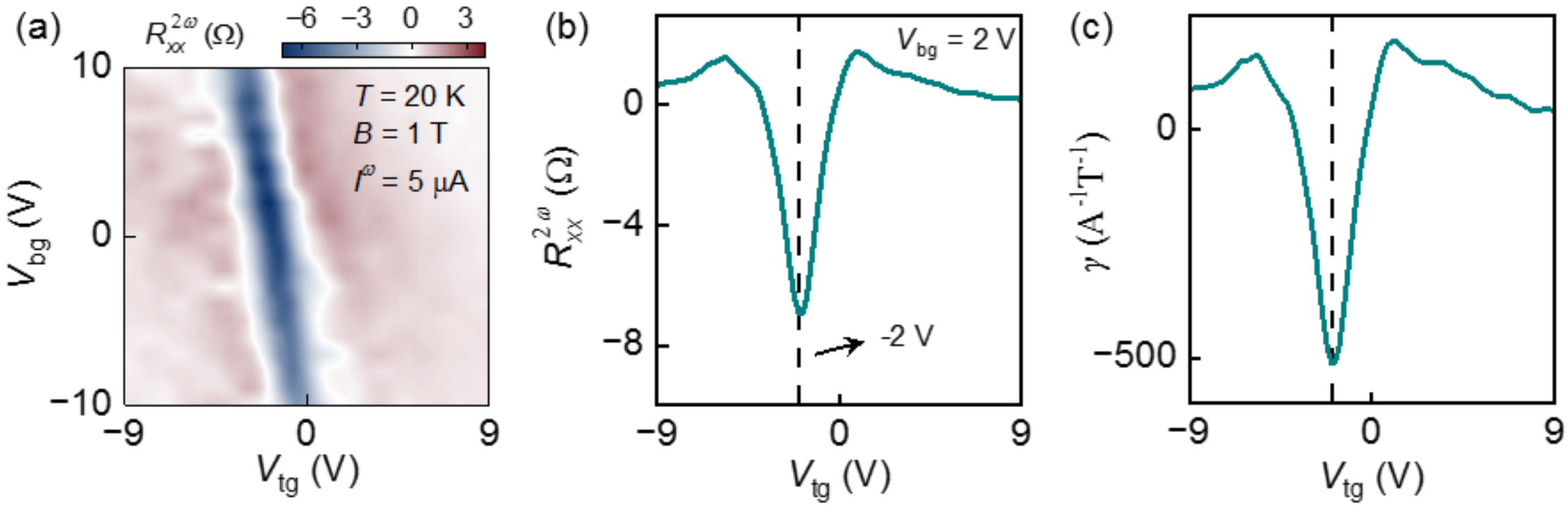

**FIG. 3**. (a) The antisymmetrized dual-gate mapping of nonreciprocal resistance $R_{xx}^{2\omega}$. (b) The top gate $V_{tg}$ dependence of the nonreciprocal resistance $R_{xx}^{2\omega}$ at a fixed back gate $V_{bg} = 2$ V. (c) The top gate $V_{tg}$ dependence of the nonreciprocal coefficient $\gamma$ under the same back gate.

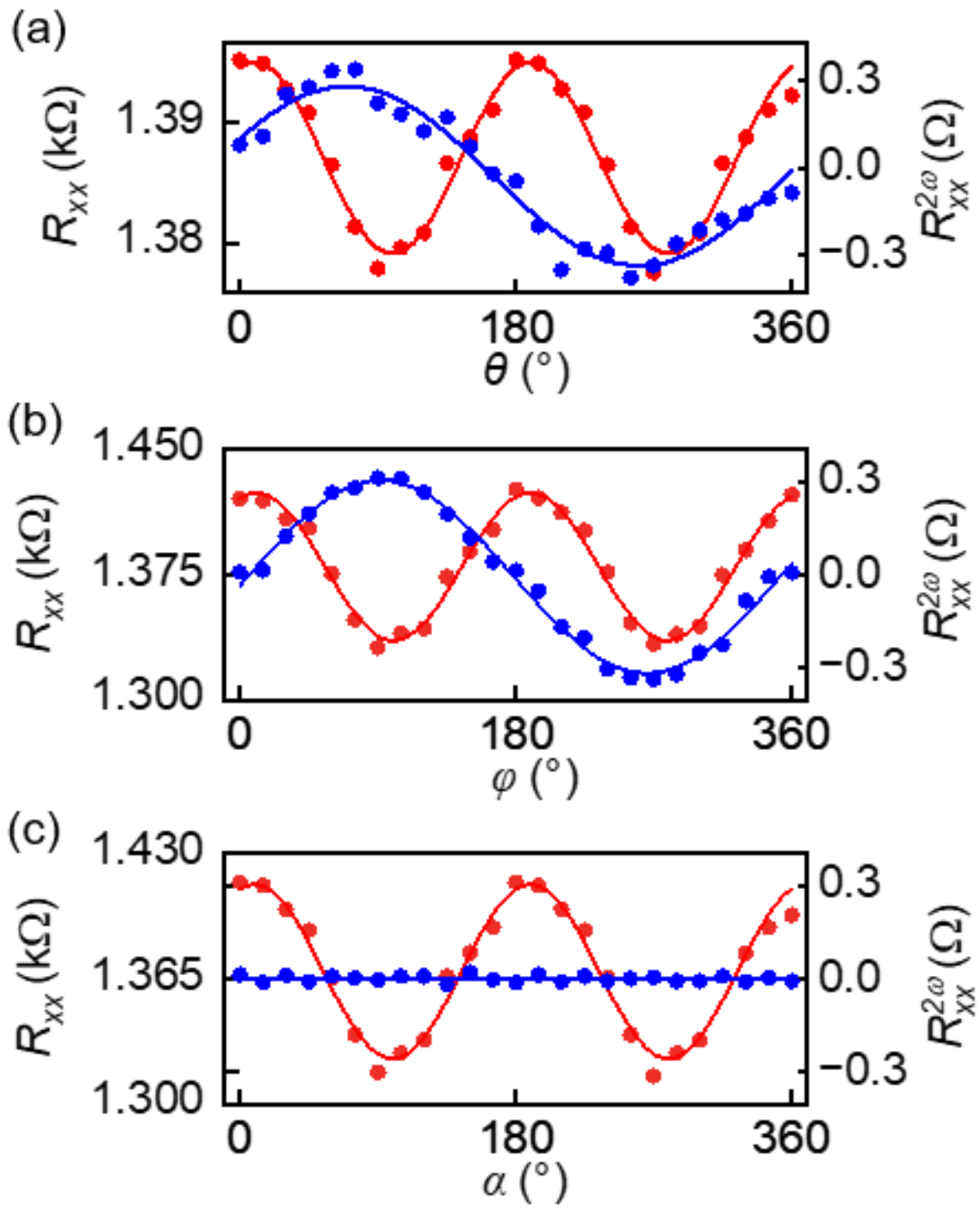

**Fig. 4.** (a) Angular dependent longitudinal resistance $R_{xx}$ (red) and nonreciprocal resistance $R_{xx}^{2\omega}$ (blue) at $T$ = 20 K and $|B|$ = 1 T with magnetic field in the *xy* plane. (b) Angular dependent behavior with magnetic field in the *yz* plane. (c) Angular dependent behavior with magnetic field in the *xz* plane, where the $R_{xx}^{2\omega}$ is nearly zero. The red lines in (a-c) and blue lines in (a, b) are fittings using trigonometric functions, while the blue line in (c) is $R_{xx}^{2\omega} = 0$.

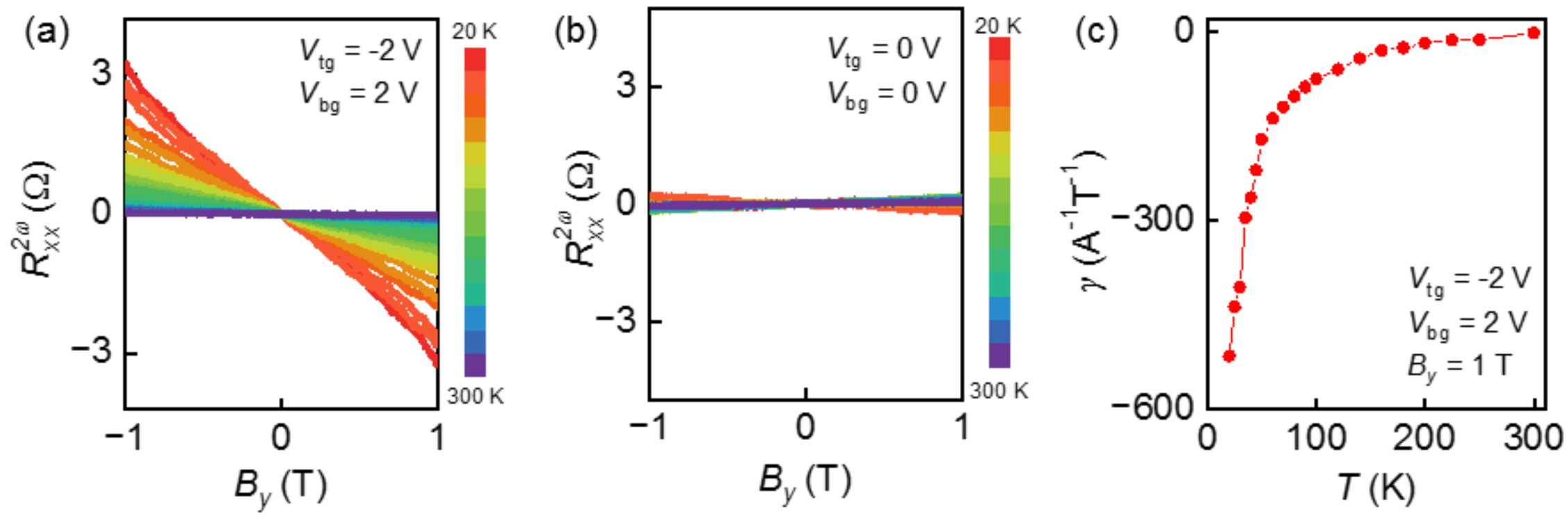

**FIG. 5.** (a) Magnetic-field dependence of the nonreciprocal resistance $R_{xx}^{2\omega}$ at different temperatures with ($V_{\rm tg}$, $V_{\rm bg}$) = (-2 V, 2 V). (b) Magnetic-field dependence of the nonreciprocal resistance $R_{xx}^{2\omega}$ at different temperatures without gate. (c) Temperature dependence of the nonreciprocal coefficient $\gamma$ at ($V_{\rm tg}$, $V_{\rm bg}$) = (-2 V, 2 V).

**Table I**. Nonreciprocal coefficient in different topological insulator materials

| Materials | $\gamma\ (A^{-1}T^{-1})$ | Reference |
|---|---|---|
| $Bi_2Se_3$ | 5 @ 60 K | Ref. 12 |
| HgTe | 400 @13 K | Ref. 11 |
| $Bi_2Se_3$/Cu(1) | 2 @4.5 K | Ref. 20 |
| $V_y(Bi_{1-x}Sb_x)_{2-y}Te_3$ | 7.4 @2.5 K | Ref. 34 |
| $BiSbTeSe_2$ | 500 @20 K | this work |